\documentclass[12pt]{article}

\newcommand{\x}{arXiv:}
\newcommand{\m}{\mathrm}
\newcommand{\be}{\begin{equation}}
\newcommand{\ee}{\end{equation}}
\newcommand{\ba}{\begin{eqnarray}}
\newcommand{\ea}{\end{eqnarray}}

\usepackage{graphicx}
\usepackage{amssymb}
\usepackage{amsmath}

\begin{document}
\thispagestyle{empty}
\begin{center}

\null \vskip-1truecm \vskip2truecm

{\Large{\bf \textsf{Holography of Little Inflation}}}

{\large{\bf \textsf{}}}

{\large{\bf \textsf{}}}

\vskip1truecm

{\large \textsf{Brett McInnes}}

\vskip1truecm

\textsf{\\  National
  University of Singapore}

\textsf{email: matmcinn@nus.edu.sg}\\

\end{center}
\vskip1truecm \centerline{\textsf{ABSTRACT}} \baselineskip=15pt
\medskip

For several crucial microseconds of its early history, the Universe consisted of a Quark-Gluon Plasma. As it cooled during this era, it traced out a trajectory in the quark matter phase diagram. The form taken by this trajectory is not known with certainty, but is of great importance: it determines, for example, whether the cosmic plasma passed through a first-order phase change during the transition to the hadron era, as has recently been suggested by advocates of the ``Little Inflation'' model. Just before this transition, the plasma was strongly coupled and therefore can be studied by holographic techniques. We show that holography imposes a strong constraint (taking the form of a bound on the baryonic chemical potential relative to the temperature) on the domain through which the cosmic plasma could pass as it cooled, with important consequences for Little Inflation. In fact, we find that holography applied to Little Inflation implies that the cosmic plasma must have passed quite close to the quark matter critical point, and might therefore have been affected by the associated fluctuation phenomena.

\newpage
\addtocounter{section}{1}
\section* {\large{\textsf{1. Holography and Hadronization in the Early Universe}}}
The description of a Quark-Gluon Plasma (QGP) is based on the quark matter phase diagram \cite{kn:ohnishi,kn:mohanty,kn:satz}, which specifies the state of the plasma in terms of the temperature $T$ and the baryonic chemical potential $\mu_B$. The plasma can take a very wide variety of different forms, ranging from the high-$T$, low-$\mu_B$ plasma explored by the ALICE experiment at the LHC (see \cite{kn:andronicover} for a recent overview with many references), to the less well-understood relatively low-$T$, high-$\mu_B$ environment being explored in the beam scan experiments at the RHIC \cite{kn:ilya,kn:dong,kn:STAR}, or to be explored at such facilities as SHINE, NICA and FAIR, and the second beam scan at the RHIC \cite{kn:shine,kn:nica,kn:fair,kn:BEAM}.

The QGP is the dominant form of matter during an important phase of the evolution of the early Universe, the plasma era which is thought to follow Inflationary reheating. It was long believed that, during this era, the only relevant region of the quark matter phase diagram is the low-$\mu_B$ region: this is understandable in view of the generally accepted value ($\eta_B \approx 10^{-9}$) of the net baryon density/entropy density ratio (which is related to $\mu_B/T$) at this point in cosmic history. Recently, however, a remarkable alternative possibility has been pointed out by Boeckel et al. \cite{kn:tillmann1,kn:tillmann2,kn:tillmann3}: it has been suggested that $\mu_B$ might in fact have been very \emph{large} (with $\mu_B/T$ ranging from unity up to $\approx 100$) during the plasma era. This is compatible with the observed baryon asymmetry, since that is generated during a short interval of ``\emph{Little Inflation}'' associated with the decay of a false QCD vacuum at the end of the plasma era.

In the conventional picture, the cosmic plasma hadronizes by passing through a smooth crossover, as is now thought to describe the QGP at low values of $\mu_B$. In the Little Inflation model, however, hadronization occurs beyond the much-discussed quark matter critical point \cite{kn:race} (believed to be located at roughly $T \approx 150$ MeV, $\mu_B \approx 150 - 300$ MeV), and therefore involves a first-order phase transition. This has many exciting consequences for the theory of primordial density fluctuations, cosmic magnetogenesis, primordial gravitational waves, and much else (for example, the very interesting ideas of Kalaydzhyan and Shuryak \cite{kn:shur} regarding the acoustics of cosmic phase transitions seem to find their most natural context in Little Inflation). In addition, the possibility of large values of $\mu_B$ during the plasma era has begun to play a role in investigations of the cosmic plasma equation of state \cite{kn:sanches}. The two alternative trajectories of the cosmic plasma in the quark matter phase diagram are shown, somewhat schematically, in Figure 1.

\begin{figure}[!h]
\centering
\includegraphics[width=0.85\textwidth]{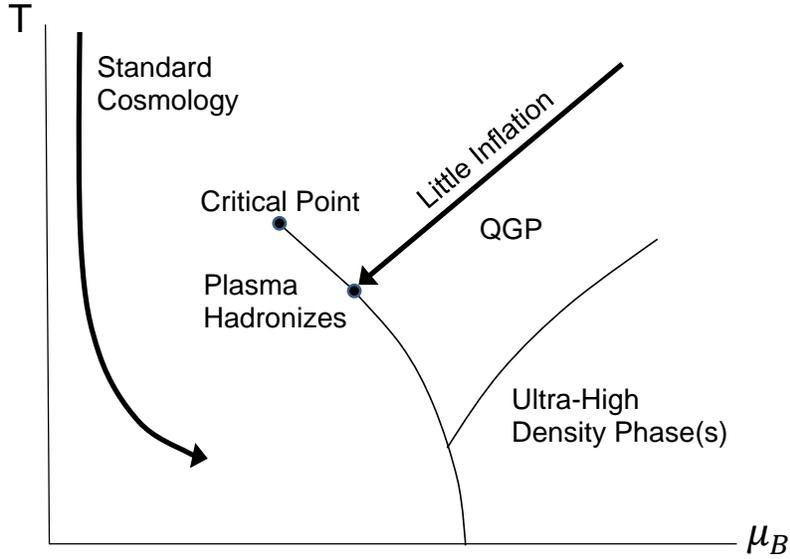}
\caption{Possible Trajectories of the cosmic plasma in the Quark Matter Phase Diagram (after Boeckel and Schaffner-Bielich \cite{kn:tillmann3}) }
\end{figure}

From the directly experimental point of view, such values of $\mu_B$ in the cosmic plasma \emph{could} mean that the high-$\mu_B$ facilities currently under construction will be the ones that will directly probe (certain aspects of) conditions in the early Universe, back to the first few microseconds; though this will only be true if the lower end of the $\mu_B/T \approx 1 - 100$ estimated range is actually realised, since those facilities are unlikely to reach very far beyond the critical point.

This region of the quark matter phase diagram is difficult to investigate theoretically. One approach \cite{kn:veron} uses the well-known ``holographic'' \emph{gauge-gravity duality}; for the specific application to heavy-ion collisions see \cite{kn:solana,kn:pedraza,kn:youngman,kn:gubser,kn:janik}. This method attempts to throw light on QCD-like thermal field theories by studying the dual, asymptotically anti-de Sitter, black hole. Here the effects of large chemical potentials (and of the strong magnetic fields arising generically in these collisions \cite{kn:skokov,kn:kharzeev1,kn:naylor}, which can strongly affect the QGP \cite{kn:kharzeev2}) can be examined by endowing the black hole with electric and magnetic charges. It is natural to ask whether this technique can be adapted to the cosmic case (where very large magnetic fields are also to be expected \cite{kn:reviewA,kn:reviewB}).

It is of course clear that both heavy ion collisions and the early Universe are very rapidly evolving systems, whereas the black hole is static. One can take the point of view that the holographic picture takes a ``snapshot'' of the system at a fixed time, but in the heavy-ion case it is also possible, though difficult \cite{kn:chesler}, to extend the theory so as to take the dynamics into account.

One can also do this in the cosmic case, though in a completely different manner that exploits the large symmetry group of FRW spacetimes \cite{kn:82}.

Consider a FRW spacetime with flat spatial sections (the only kind we shall consider here); we can construct it in the following manner. Let $\psi,\;\zeta,\;\xi$ be dimensionless coordinates on a flat three-dimensional space, and consider the corresponding flat spacetime with metric $-\text{d}t^2\;+\;L^2\left[\text{d}\psi^2\;+\;\text{d}\zeta^2\;+\;\text{d}\xi^2\right]$, where $L$ is some parameter with dimensions of length. Let $a(t)$ be any smooth function of $t$, and define $\tau$ by $\text{d}\tau = a(t)\text{d}t$. Then a conformal transformation of the flat spacetime, with conformal factor $a(t)^2$, produces the FRW metric $-\text{d}\tau^2\;+\;a(t)^2L^2\left[\text{d}\psi^2\;+\;\text{d}\zeta^2\;+\;\text{d}\xi^2\right]$ (where now $t$ and $a(t)$ are to be interpreted as certain functions of $\tau$). Notice that the full spacetime, unlike its spatial sections, is not in general flat; but it is \emph{conformally} flat.

Let us arbitrarily fix a two-dimensional flat surface $\mathcal{S}$ in a three-dimensional spatial slice of the FRW spacetime. We can orient our coordinates so that $\xi$ is perpendicular to $\mathcal{S}$, and so that $\psi,\;\zeta$ are coordinates in it. Then $\mathcal{S}$ can be regarded as any of the surfaces $\xi = $ constant, and, by adjoining the time coordinate, one can define a \emph{three-dimensional sub-spacetime} $\mathcal{S}^t$ (signature $1+2$) embedded in the four-dimensional FRW spacetime, with spacetime metric $-\text{d}\tau^2\;+\;a(t)^2L^2\left[\text{d}\psi^2\;+\;\text{d}\zeta^2\right]$. This three-dimensional sub-spacetime is clearly still conformally flat, with transformed metric\footnote{See the discussion immediately after equation (\ref{A}) below.} $-\text{d}t^2\;+\;L^2\left[\text{d}\psi^2\;+\;\text{d}\zeta^2\right].$

Our strategy now is as follows. In most spacetimes, one would not expect to obtain a satisfactory description by restricting attention to three-dimensional sub-spacetimes like those we have been discussing. \emph{But FRW spacetimes are very special}: by construction, they are homogeneous and isotropic at each point. It follows that, for FRW spacetimes, the physics of any given three-dimensional sub-spacetime like $\mathcal{S}^t$ dictates the physics of the full four-dimensional spacetime. This point of view is actually the most natural one when we are studying cosmic magnetic fields, because the latter are always associated with a \emph{flux} through some (compact domain in a) two-dimensional surface, and indeed homogeneity and isotropy ensure that the field is fully specified if we know this flux through one such surface. More generally, if we have a FRW spacetime containing a plasma of temperature $T$ and baryonic chemical potential $\mu_B$, and a magnetic field of strength $B$, we claim that we can understand the relations between these quantities \emph{if we can understand them when they are restricted to $\mathcal{S}^t$.} In a sense, the extremely large group of symmetries of the FRW spacetime allows us to regard it as being ``effectively three-dimensional''.

The idea now is to regard $\mathcal{S}^t$ (which is not flat, but which \emph{is} conformally flat) as the (three-dimensional) conformal boundary of a four-dimensional asymptotically AdS bulk spacetime, and then to use the bulk physics to constrain the bulk quantities corresponding to $T, \mu_B,$ and $B$. Holography then translates these constraints back to $\mathcal{S}^t$, and then they can be extended to the full four-dimensional FRW spacetime. The process might be symbolized as ``$4\;\rightarrow 3 \; \rightarrow 4$''.

The reader may object that the bulk spacetime is static, so the bulk quantities corresponding to $T, \mu_B,$ and $B$ do not evolve: how then can holography constrain these parameters? Consider again the magnetic flux through some domain in a two-dimensional plane $\mathcal{S}$ in a spatial section of the FRW spacetime. This flux has a remarkable property: it is \emph{conformally invariant} with respect to the transformations we discussed above. For, at least in conventional cosmology ---$\,$ see below ---$\,$ the magnetic field dilutes with the cosmic expansion at precisely the inverse of the rate at which the area of the two-dimensional surface is stretched (that is, $B \propto a(t)^{-2}$). The flux does not evolve: it does not ``know'' whether it is being evaluated on $\mathcal{S}^t$ with the metric $-\text{d}\tau^2\;+\;a(t)^2L^2\left[\text{d}\psi^2\;+\;\text{d}\zeta^2\right]$ or with the conformally transformed metric $-\text{d}t^2\;+\;L^2\left[\text{d}\psi^2\;+\;\text{d}\zeta^2\right].$

Now in fact this comment applies to several other interesting quantities: we will see that combinations like $B/T^2$ and $\mu_B/T$ are likewise invariant with respect to the conformal transformation with conformal factor $a(t)^2$ ---$\,$ in other words, they are \emph{constant} as the cosmic plasma evolves. For example, $T \propto a(t)^{-1}$, but similarly $\mu_B \propto a(t)^{-1}$. If we can use holography to constrain these \emph{ratios}, then the conformal transformation that, as above, restores the time dependence, will have no effect on such constraints. Thus, finally, we obtain constraints on the time-dependent plasma in the full four-dimensional FRW spacetime.

This multi-step approach to cosmic holography is very limited: it only works for (spatially flat) FRW spacetimes, and it only allows us to study a small range of physically interesting ratios, those which are conformally invariant in the cosmic sense (that is, constant with respect to cosmic time). Nevertheless it will prove to be useful.

The bulk spacetime must be foliated by three-dimensional sections (of signature $1+2$) transverse to the radial direction (that is, they correspond to $r =$ constant), so it is an asymptotically AdS spacetime containing an (electrically and magnetically) charged black hole with a \emph{planar} event horizon, sometimes called a ``black brane''. In \cite{kn:82} we
studied such spacetimes from the point of view of string theory; specifically, we asked whether it was consistent to assume that string-theoretic objects, such as branes, can always be neglected in the bulk, even under optimal conditions (the string coupling and the ratio of the string length scale to the AdS curvature scale $L$ are small). We found that this is \emph{not} the case, because under some circumstances the black hole itself begins to generate branes and radiate them towards infinity. Requiring that this instability should not arise imposes a bound on the conformally invariant ratio $B/T^2$: one speaks of a \emph{holographic bound on the cosmic magnetic field} during the plasma era. It transpires that this bound is in fact satisfied, though not by a large margin, in the current models of cosmic magnetogenesis.

However, in \cite{kn:82} we followed the standard assumption that the cosmic baryonic chemical potential is negligible throughout the plasma era, and so we have to revise those results in the light of a possible episode of Little Inflation at the end of the plasma era. One can see that there is an issue here, because a non-negligible chemical potential corresponds to a non-zero electric charge on the dual black hole. Since the electric and magnetic charges enter symmetrically into the black hole metric (as a result of the classical electromagnetic duality of Maxwell's equations), the chemical potential has a similar effect to a large magnetic field, and likewise threatens to trigger a ``stringy'' instability. This constrains Little Inflation by bounding the (conformally invariant) ratio of the baryonic chemical potential to the temperature.

In this work we extend the methods of \cite{kn:82} to study this holographic constraint. It proves to be very stringent: the range $\mu_B/T \; \approx \; 1 - 100$ discussed in \cite{kn:tillmann3} is greatly narrowed to $\approx 1\; \leq \; \mu_B/T \; \leq \;\; \approx 2.35$. Since the value of $\mu_B/T$ at the quark matter critical point provides the lower bound around 1, this result means that, if the cosmic plasma does indeed undergo a first-order phase transition at the end of the plasma era, it must pass close to the critical point. Thus the ``race'' to locate that point, and to identify the physics associated with it \cite{kn:race}, assumes cosmological significance\footnote{One should however be aware that the cosmic plasma differs in some ways from the plasma produced in heavy ion collisions: see below for a detailed discussion of this. On the other hand, certain properties of the QGP will be important in both cases.}.

In short, if Little Inflation is to be compatible with holography, then it must occur in precisely that part of the quark matter phase diagram where remarkable phenomena associated with the quark matter critical point may soon be observed, but \emph{also} where a holographic instability is not far off.

We begin with a brief review of the relevant bulk geometry and of its holographic interpretation in this application.

\addtocounter{section}{1}
\section* {\large{\textsf{2. Planar AdS Black Holes and FRW Holography}}}
The bulk geometry is described by a ``Charged Planar AdS Black Hole'' metric \cite{kn:lemmo}\cite{kn:dyon}, a solution of the AdS Einstein-Maxwell system\footnote{Note that apart from the trivial case with $Q^* = P^* = 0$, none of these metrics is an Einstein metric. The effect with which we will be concerned below, in Section 3, has mostly been studied in the (Euclidean) Einstein case; see for example \cite{kn:ferrari} and references therein.} given by
\begin{eqnarray}\label{A}
g(\m{CPAdSBH}) & = & -\, \Bigg[{r^2\over L^2}\;-\;{8\pi M^*\over r}+{4\pi (Q^{*2}+P^{*2})\over r^2}\Bigg]\text{d}t^2\; \nonumber \\
& &  + \;{\text{d}r^2\over {r^2\over L^2}\;-\;{8\pi M^*\over r}+{4\pi (Q^{*2}+P^{*2})\over r^2}} \;+\;r^2\Big[\text{d}\psi^2\;+\;\text{d}\zeta^2\Big].
\end{eqnarray}
Here $\psi$ and $\zeta$ are dimensionless coordinates on the planar sections transverse to the radial coordinate $r$, $L$ is the asymptotic AdS curvature radius, and $M^*$, $Q^*$, and $P^*$ are geometric parameters related respectively to the mass, electric charge, and magnetic charge per unit horizon area. (See \cite{kn:82,kn:77} for the details.) \emph{Notice that the (conformal) metric at infinity for this spacetime is represented by the metric $-\text{d}t^2\;+\;L^2\left[\text{d}\psi^2\;+\;\text{d}\zeta^2\right]$ on $\mathcal{S}^t$ in our earlier discussion.} We can think of the transverse sections $r$ = constant as deformed copies of a sub-spacetime of a FRW spacetime, as explained above.

In the usual way $M^*$, $Q^*$, and $P^*$ determine (for a fixed value of $L$) the value of $r$ at the event horizon, $r = r_h$: we have
\begin{equation}\label{B}
{r_h^2\over L^2}\;-\;{8\pi M^*\over r_h}+{4\pi (Q^{*2}+P^{*2})\over r_h^2} = 0.
\end{equation}

The potential for the electromagnetic field outside the black hole is
\begin{equation}\label{C}
A\;=\;\left ({1\over r_h}\,-\,{1\over r}\right ){Q^*\over L}\text{d}t\;+\;{P^*\over L}\psi \text{d}\zeta,
\end{equation}
where the constant term in the coefficient of $\text{d}t$ ensures that the Euclidean version of this one-form is regular. The field strength is
\begin{equation}\label{D}
F\;=\;-\,{Q^*\over r^2L}\text{d}t \wedge \text{d}r \;+\;{P^*\over L}\text{d}\psi \wedge \text{d}\zeta.
\end{equation}

Now we turn to the dual field theory on the boundary. The quark chemical potential of this system is related holographically to the asymptotic value of the time component of the potential form, while the magnetic field $B$ of the dual system is related to the asymptotic value of the field strength \cite{kn:klebwit,kn:hartkov,kn:koba}. We therefore have, using the customary baryonic chemical potential $\mu_B$, and compensating for the fact that $\text{d}\psi$ and $\text{d}\zeta$ are dimensionless,
\begin{equation}\label{E}
\mu_B\;=\;{3Q^*\over r_hL}
\end{equation}
and
\begin{equation}\label{F}
B \;=\; P^*/L^3.
\end{equation}
The temperature of the boundary system is that of the Hawking radiation of the black hole, which is given by
\begin{equation}\label{G}
T\;=\;{1\over 4\pi r_h}\,\Bigg({3r_h^2\over L^2}\;-\; {4\pi (Q^{*2}+P^{*2}) \over r_h^2}\Bigg),
\end{equation}
where we have used equation (\ref{B}).

Combining equations (\ref{E}), (\ref{F}), and (\ref{G}), we obtain
\begin{equation}\label{H}
3r_h^4\,-\,4\pi TL^2r_h^3\,-\,{4\pi \over 9}\mu_B^2L^4r_h^2\,-\,4\pi B^2L^8\;=\;0.
\end{equation}
If the temperature is positive, then an event horizon exists and so this quartic can be solved for $r_h$, which can then be regarded as a function of the boundary parameters $T$, $\mu_B$, and $B$. In fact, given $L$ and these three quantities, $r_h$ can be computed in this manner, and then the black hole parameters $M^*$, $Q^*$, and $P^*$ can be reconstructed from equations (\ref{B}), (\ref{E}), and (\ref{F}).

These last three quantities are of course constants, both in the bulk and in the obvious (flat) boundary geometry. In the cosmological application, all of them must be promoted to functions of \emph{cosmic} time, since the dual quantities $T,$ $\mu_B,$ and $B$ are such functions; but from the bulk point of view, cosmic time is not a time coordinate but rather a parameter along a curve in the abstract three-dimensional space of planar AdS black hole metrics given in equation (\ref{A}). We will see that the three ``coordinates'' ($M^*$, $Q^*$, $P^*$) depend on this parameter through the FRW scale factor $a(t)$; this makes it straightforward to focus on conformally invariant quantities, which can be regarded as being defined on the flat spacetime to which the FRW spacetime is conformally related, as explained in the preceding Section. In detail, this works as follows.

First, for a plasma, $T$ decreases according to $1/a(t)$. In a simple Boltzmann model (like the one used in \cite{kn:phobos}), the antimatter/matter ratio is given by exp$(-\,2\,\mu_B/T)$, so, in any regime in which this ratio does not change rapidly, $\mu_B$ likewise decreases in accordance with $1/a(t)$: \emph{$\mu_B/T$ is a conformal invariant}. (As the temperature drops, massive particles annihilate and their entropy is transferred to effectively massless particles, which implies that this naive model of the particle populations can only be approximate. This approximation is nevertheless adequate for our purposes; one might wish to apply it only to the plasma immediately prior to the phase change.) The trajectory in the quark matter phase diagram is therefore straight (see Figure 2 in \cite{kn:tillmann3} and Figure 1 above), and we have
\begin{equation}\label{I}
\mu_B = \varsigma_B \, T,
\end{equation}
where $\varsigma_B$, the ``specific baryonic chemical potential'', is a positive constant, the reciprocal of the slope of the straight line\footnote{Note that, because we are (for simplicity) \emph{not} compactifying the planar sections here, there is no Hawking-Page transition for these black holes \cite{kn:surya}, so we need not be concerned that such a transition will interfere before the dual plasma hadronizes. The Hawking-Page transition can be restored, at any desired temperature, by compactifying the planar event horizon to a torus \cite{kn:AdSRN}; in our case it would be natural to choose it to occur at the temperature at which the cosmic plasma crosses the phase line.}. Our principal objective in this work is in fact to constrain $\varsigma_B$, by regarding it as a conformal invariant, in the sense discussed earlier.

Similarly, in conventional cosmology\footnote{Alternative possible evolution laws for $B$ have been proposed \cite{kn:barrow1,kn:barrow2,kn:barrow3}, but are controversial \cite{kn:durrer,kn:sahni,kn:cost}; if they can arise, they can probably only do so \emph{before} the plasma era we are studying here \cite{kn:kandus}. During the plasma era, such ``superadiabatic amplification'' can be reconciled with a holographic bound \cite{kn:82} only with difficulty; see below.}, the magnetic field decreases according to $1/a(t)^2$, so $B/T^2$ is a conformal invariant. Again, $B/T^2$ is the kind of ratio which we can hope to constrain by means of holography, and that was done (when $\varsigma_B = 0$) in \cite{kn:82}.

Now regard equation (\ref{H}) as the \emph{definition} of $r_h$, which now becomes a function of cosmic time in the FRW spacetime: that is, it is defined to be the (largest) solution of this equation, given the coefficient functions $T$, $\mu_B$, and $B$. As the solution of a quartic equation, it depends on these functions in a very complicated way. Remarkably enough, however, its evolution with cosmic time is extremely simple: by inspecting equation (\ref{H}), given the above evolution laws for $T$, $\mu_B$, and $B$, one sees that $r_h$ decreases according to $1/a(t)$. In the cosmological case one can then define $Q^*$ by equation (\ref{E}), and $P^*$ by equation (\ref{F}); because of the evolution law for $r_h$, one finds that both evolve in the same way (as should be the case, according to electromagnetic duality\footnote{This would \emph{not} be the case if $B$ evolved non-adiabatically, because then $r_h$ would evolve in a much more complicated way.}), namely with $1/a(t)^2$. Finally, equation (\ref{B}) defines $M^*$ for the FRW spacetime, and shows that it evolves according to $1/a(t)^3$.

The important consequence of all this is that equations (\ref{B}), (\ref{E}), (\ref{F}), and (\ref{H}) can all be interpreted either in the black hole bulk, \emph{or} (by holography) in the flat space dual field theory, \emph{or} (by multiplying both sides of the equation by a suitable power of the scale factor $a(t)$) in the three-dimensional expanding sub-spacetime, $\mathcal{S}^t$. This is certainly not a trivial statement: it is a consequence of the fact that the geometry is ``assembled'' from components which are fundamentally \emph{planar}. For example, if we had used an asymptotically AdS black hole with a \emph{spherical} event horizon, then equation (\ref{B}) would have taken the form
\begin{equation}\label{J}
{r_h^2\over L^2}\;+\;1\;-\;{2 M\over r_h}+{Q^2+P^2\over 4\pi r_h^2} = 0,
\end{equation}
where $M$, $P$, and $Q$ are the usual (finite) mass and charge parameters; but clearly this equation \emph{cannot} transform in a homogeneous way under conformal transformations. The formula for the Hawking temperature likewise acquires terms that rule out the above procedure: it is unique to the planar case.

In the conventional picture of the evolution of the cosmic plasma, the specific baryonic chemical potential $\varsigma_B$ (equation (\ref{I})) is extremely small; whereas in Little Inflation it is large, potentially as large as 100. Thus $\varsigma_B$ is the central object of attention here, and the sequel is devoted to explaining how holography constrains it.

\addtocounter{section}{1}
\section* {\large{\textsf{3. The Brane Action}}}
Our approach to FRW spacetimes focuses on two-dimensional planes embedded in the spatial sections, and on the associated three-dimensional spacetimes $\mathcal{S}^t$. Each transverse section $r$ = constant in the bulk spacetime with metric given in equation (\ref{A}) is a deformed copy of $\mathcal{S}^t$, and so it is natural to investigate the behaviour of these copies in the bulk geometry.

In \cite{kn:82} we argued that these transverse sections can be studied by a simple function $\mathfrak{S}(r)$ defined (for four-dimensional asymptotically AdS black hole spacetimes with planar sections) in the following manner. Let $A_r$ be the Lorentzian area of an (arbitrarily chosen) compact domain\footnote{The reader may prefer to transfer this discussion to the Euclidean domain, in which $t$ is compactified, and the ``planar'' coordinates $\psi$ and $\zeta$ are naturally converted to coordinates on a torus. Then ``area'' and ``volume'' have their conventional connotations and are automatically finite. The final answer can then be straightforwardly continued back to the Lorentzian domain.} in the three-dimensional section (including the time axis) located at $r$, and let $V_r$ denote the Lorentzian volume of the four-dimensional bulk region between the event horizon and that domain. Then we define
\begin{equation}\label{K}
\mathfrak{S}(r)\; \equiv \; A_r\;-\;{3\over L}\,V_r,
\end{equation}
with the understanding that this quantity is defined only up to an overall positive multiplicative constant, which we shall choose so that $\mathfrak{S}(r)$ is dimensionless.

For planar submanifolds of AdS$_4$ itself (regarded as the $r_h \rightarrow 0$ limit of the black hole), $\mathfrak{S}(r)$ vanishes identically; but that is not so for AdS black hole spacetimes, which are merely \emph{asymptotically} AdS. In that case, $\mathfrak{S}(r)$ vanishes at the event horizon\footnote{The Lorentzian area of the event horizon, including the time direction, is zero, since it is a null surface; and of course $V_r$ also vanishes there, by its definition. One sees this more clearly in the Euclidean version, where the event horizon becomes the origin of a polar coordinate system.}, and it is always positive nearby. Far from the event horizon, however, the situation is less clear, since it is characteristic of asymptotically AdS geometries that areas and volumes grow at much the same rate. In fact, $\mathfrak{S}(r)$ can even become negative far from the event horizon: eventually the volume can overcome the area.

That does not happen for the planar AdS-Schwarzschild geometry (see \cite{kn:82}); nor does it happen for the charged planar AdS black holes studied in the preceding Section, as long as the charges are fairly small. But, as we shall see, it \emph{can} happen if the charges are large, yet still sub-extremal\footnote{The black hole with metric (\ref{A}) has extremal or sub-extremal charges if equation (\ref{B}) has a positive real solution: the condition for that is $\left (P^{*2}+Q^{*2}\right )^3 \leq (27/4)\pi M^{*4}L^2$.}.

As we explained in \cite{kn:82}, allowing $\mathfrak{S}(r)$ to become negative, that is, smaller than its value at the event horizon, has serious consequences. For it was shown by Seiberg and Witten \cite{kn:seiberg} (see also \cite{kn:wittenyau}), that $\mathfrak{S}(r)$ is, up to a positive multiplicative factor proportional to the tension of the brane, nothing but \emph{the action of a BPS 2-brane} wrapping around $r = $ constant. Branes nucleating near to the event horizon, where this action is positive, will tend to contract back into the event horizon, where the action vanishes. If however there is a region beyond some value of $r$ in which this action is lower than it is in the vicinity of the event horizon, then a brane nucleating in that region will tend to escape to infinity instead of contracting back into the black hole, and the system becomes unstable. The dual phenomenon in the field theory is that a certain scalar field, even if suppressed initially (on the grounds that there is no such field in QCD), begins to grow and quickly dominates the gauge fields. Various aspects of such phenomena have been discussed recently in \cite{kn:ferrari} and \cite{kn:maloney}.

We can evaluate $\mathfrak{S}(r)$ for the metric in equation (\ref{A}): it is
\begin{equation}\label{L}
\mathfrak{S}_{CPAdSBH}(r)\;=\; {r^2\over L^2}\sqrt{{r^2\over L^2}-{8\pi M^*\over r}+{4\pi (P^{*2}+Q^{*2})\over r^2}}-{r^3\over L^3}+{r_h^3\over L^3},
\end{equation}
where the last two terms correspond to the volume term in equation (\ref{K}). This may be written more usefully as
\begin{equation}\label{M}
\mathfrak{S}_{CPAdSBH}(r)\;=\; { \left (-8\pi M^* + {4\pi (P^{*2}+Q^{*2})\over r} \right )/L\over 1+\sqrt{1-{8\pi M^*L^2\over r^3}+{4\pi (P^{*2}+Q^{*2})L^2 \over r^4}}}+{r_h^3\over L^3}.
\end{equation}
This function is non-negative if and only if its value as $r \rightarrow \infty$ is non-negative: that is, we need
\begin{equation}\label{N}
-\,4\pi M^* L^2 + r_h^3 \;\geq \;0
\end{equation}
if the bulk is to be stable. Notice that this inequality is well-defined, in the sense that both terms on the left evolve according to $a(t)^{-3}$ when we transfer to the cosmological spacetime. (The reader can verify that analogous statements hold for all of our subsequent equations and inequalities.)

We conclude that the requirement that the holographic picture should be internally consistent imposes a constraint on the black hole parameters in the bulk. Our next task is to determine what this means for the boundary theory and the conformally related FRW spacetime.

\addtocounter{section}{1}
\section* {\large{\textsf{4. The Bound on $\mu_B/T$}}}
Using equation (\ref{B}), we can write (\ref{N}) as
\begin{equation}\label{O}
4\pi (P^{*2}+Q^{*2})L^2 \;\leq \;r_h^4.
\end{equation}
Inserting this into equation (\ref{G}) we obtain
\begin{equation}\label{P}
2\pi TL^2 \geq r_h.
\end{equation}
Combining (\ref{O}) and (\ref{P}) with equations (\ref{E}) and (\ref{F}), we obtain the fundamental inequality\footnote{In terms of the black hole parameters, the inequality (\ref{Q}) is expressed as $\left (P^{*2}+Q^{*2}\right )^3 \leq 4\pi M^{*4}L^2$. Censorship (which we found earlier to demand $\left (P^{*2}+Q^{*2}\right )^3 \leq (27/4)\pi M^{*4}L^2$) is therefore always ensured here, though not by a very large margin.}
\begin{equation}\label{Q}
B^2\;+\;{\mu_B^2r_h^2\over L^4}\;\leq \;4\pi^3T^4.
\end{equation}
Thus we see that holography imposes a bound, given the temperature, on this combination of the magnetic field and the baryonic chemical potential. Bear in mind, however, that $r_h$ is to be regarded (via equation (\ref{H})) as a function of $B$, $T$, and $\mu_B$, obtained by solving a quartic equation; so, expressed in terms of the physical parameters, this relation is actually very complex. Furthermore, it involves $L$, which is not fixed in any obvious way here: it would be preferable if our final conclusions were independent of that quantity. So we need to examine (\ref{Q}) more carefully.

Our specific objective in this work is to constrain the physical parameters of the cosmic plasma at the time when it hadronizes. As we know, there are two proposals for the manner in which this happens: fortunately, both of them correspond to particularly simple special cases of the inequality (\ref{Q}).

$\bullet$ In the conventional picture of the evolution of the cosmic plasma, the trajectory in the quark matter phase plane is very close to the $T$ axis, so that the cosmic plasma passes through a smooth crossover on its route to hadronization: there is no first-order phase transition and no Little Inflation. In that picture, then, $\mu_B$ is negligible throughout the plasma era, and (\ref{Q}) reduces to
\begin{equation}\label{R}
B\;\leq \;2\pi^{3/2}T^2.
\end{equation}
This is the bound on cosmic magnetic fields explained in \cite{kn:82}; it implies a bound of $\approx \; 3.6 \times 10^{18}$ gauss at the hadronization temperature. In this picture, cosmic magnetogenesis may be associated with Inflation (see for example \cite{kn:tasinato}\cite{kn:bamba}), and the magnetic field energy densities involved can be enormous, up to equipartition with the plasma density; so a bound on $B$ is of interest. Furthermore, this bound is important because it very strongly constrains unconventional evolution laws for $B$, such as the one discussed in \cite{kn:cost}; for, in that case, $B/T^2$ would no longer be constant but would grow by a very large factor (depending on the reheating temperature) during the plasma era, so it becomes difficult to satisfy a bound like (\ref{R}) at all times.

$\bullet$ In Little Inflation, $\mu_B$ is far from negligible, so that the cosmic plasma does pass through a first-order phase transition. But while this theory may possibly give a viable account of cosmic magnetogenesis (see the discussion around Figure 18 in \cite{kn:reviewB}), the magnetic fields involved are \emph{relatively} small, about $10^{-4}$ times the values typical of inflationary magnetogenesis. (The magnetic field is generated along with the baryon asymmetry, so the magnetic energy density is in the vicinity of equipartition with the baryonic, rather than the plasma, energy density.) One can therefore assume that $B$ is negligible for our purposes\footnote{However, in view of the uncertainties currently attending all theories of cosmic magnetogenesis, one should consider the possibility that magnetic fields are larger in Little Inflation than expected ---$\,$ for example, relics of inflationary magnetogenesis might be important. If that were the case, the effect would be to \emph{strengthen} our conclusions, in the sense that a detailed analysis of (\ref{Q}) shows that the upper bound on $\mu_B/T$ we are about to deduce would be lowered ---$\,$ though only to a small extent, even for very large fields. These facts are discussed in the Appendix to this paper.}, and this greatly simplifies the situation because equation (\ref{H}) is now quadratic rather than quartic.

Before we proceed to the solution, we should stress that ignoring $B$ would usually be a very poor approximation in the case of the plasma produced in a heavy ion collision \cite{kn:skokov,kn:kharzeev1,kn:naylor,kn:kharzeev2}. Furthermore, we are ignoring the effects of cosmic \emph{vorticity}: that is the customary assumption (though it might be desirable under some circumstances to reconsider it \cite{kn:brand}); but, again, the analogous assumption, that the angular momentum density is negligible, is certainly not normally justified in the heavy-ion case, where the holographic dual is a black hole endowed with angular momentum, as in \cite{kn:klemm,kn:77,kn:shear,kn:79}. Again, as we have seen, the time evolution of all physical parameters in the cosmic case is controlled in a simple way by a single function, $a(t)$; but it is not clear that any such simple description of the dynamics is possible for a heavy-ion plasma. Finally, the time scales in the two cases are very different: the heavy ion plasma exists for a time typical of strong-interaction physics (measured in femtometres/c), while the cosmic plasma endures for several microseconds. This is a crucial distinction for any discussion based, as ours is here, on the development of an instability. Thus, our results do not extrapolate to the heavy-ion case in any straightforward way. However, those features of the QGP (most importantly, its behaviour near to the critical point) which are independent of the dynamics will be common to both kinds of plasma.

Now solving (\ref{H}) we have
\begin{equation}\label{S}
r_h \;=\;{2L^2\over 3}\left (\pi T + \sqrt{\pi^2T^2+(\pi /3)\mu_B^2}\right ).
\end{equation}
It is clear that when (\ref{S}) is substituted into (\ref{Q}), $L$ drops out, and, in the absence of $B$, $\mu_B$ and $T$ are the only quantities remaining. Since they are proportional to each other, (\ref{Q}) can in this case be reduced to an expression involving $\varsigma_B$ (equation (\ref{I})) only: we find
\begin{equation}\label{T}
\pi \varsigma_B\;+\;\varsigma_B \sqrt{\pi^2+(\pi /3)\varsigma_B^2}\;\leq \;3\pi^{3/2}.
\end{equation}
Some algebra simplifies this to
\begin{equation}\label{U}
\varsigma_B^4\;+\;18\pi^{3/2}\varsigma_B \;-\;27\pi^2 \;\leq\; 0.
\end{equation}
The quartic here is strictly increasing for $\varsigma_B \geq 0$, so its sole positive root yields an upper bound on $\varsigma_B$. This root is exactly $\left(1\;-\;2^{1/3}\;+\;2^{2/3} \right)\sqrt{\pi}$, and so we have finally, restoring $\mu_B$ and $T$,
\begin{equation}\label{V}
\mu_B/T \; \leq \; \left(1\;-\;2^{1/3}\;+\;2^{2/3} \right)\sqrt{\pi}\;\approx 2.353.
\end{equation}

The key datum now is the location of the quark matter critical point. To see why this is so, refer to Figure 1: the phase transition line slopes \emph{downwards} from the critical point, into regions of larger $\mu_B$ but smaller $T$. Therefore, if the trajectory of the cosmic plasma in the phase diagram intercepts the transition line away from the critical point, it does so at larger values of $\mu_B/T$ than the value at the critical point: in short, the value at the critical point puts a \emph{lower} bound on $\mu_B/T$ at the point where the cosmic plasma hadronizes. Thus, Little Inflation itself, combined with holography, allows us to constrain $\mu_B/T$ from both sides.

Now in fact the precise location of the critical point is a matter of intense interest \cite{kn:race}, and there is reason to hope that it will be settled in the near future. Theoretical estimates, for example from lattice theory, have become considerably more precise in recent years \cite{kn:lattice}. (There is a growing consensus that the critical temperature is around 150 MeV; it is the critical value of the baryonic chemical potential that is most difficult to compute.) It is interesting to note that in the past (see for example \cite{kn:mohantyold}), lattice-theoretical estimates of the critical value of $\mu_B$ were in the 350-450 MeV range, threatening a conflict with our inequality (\ref{V}); but, more recently \cite{kn:race}, a value for $\mu_B/T$ at the critical point around $1 - \,2$ has come to be favoured. More recently still, however, a sigma-model approximation approach \cite{kn:ayalasigma} has indicated that higher values may be possible, while an analysis of the most recent experimental data apparently suggests a value \emph{below} unity \cite{kn:lacey}.

To be definite, let us settle on the range given in \cite{kn:race}; then we can summarize the situation by stating that the holographic version of Little Inflation requires that, for an interval of time\footnote{Following \cite{kn:tillmann3}, and as discussed above, we are assuming here that $\mu_B/T$ is constant during this time. Note also that the holographic picture of the plasma only describes it when it is strongly coupled, which may only have been the case during the late plasma era; so we do not claim that our bound applies at all times.} immediately before the cosmic plasma underwent a first-order phase transition to the hadronic state, $\mu_B/T$ must have satisfied
\begin{equation}\label{W}
\approx 1\; \leq \; \mu_B/T \; \leq \;\; \approx 2.35.
\end{equation}
This is indeed a remarkable refinement of the range given in \cite{kn:tillmann3}, $1 - 100$.

If, for example, we put the critical point at $\mu_B$ = 300 MeV, $T$ = 150 MeV, and assume for definiteness that the transition line near to the critical point makes an angle of (very roughly) 45 degrees with the horizontal, then a simple calculation shows that Little Inflation can be compatible with holography only if the cosmic plasma hadronizes between $T  \approx 140 - 150$ MeV, $\mu_B  \approx 300 - 315$ MeV. This is very interesting, for two reasons. First, it means that the cosmic plasma hadronizes at a point well within the range probably accessible to near-future facilities such as SHINE, NICA, FAIR, and the upgraded RHIC \cite{kn:shine,kn:nica,kn:fair,kn:BEAM}. Second, it means that the trajectory of the cosmic plasma in the quark matter phase diagram must have passed very near to the critical point itself; this means that the plasma might possibly have experienced the characteristic fluctuation phenomena associated with critical points, such as the QCD version of \emph{critical opalescence} \cite{kn:csorgo}. That could have all manner of important consequences.

\addtocounter{section}{1}
\section* {\large{\textsf{5. Conclusion: Constraining Cosmic Hadronization}}}
Little Inflation presents a version of cosmic history which differs very distinctly from the conventional picture. Perhaps its most exciting feature is that it brings the cosmology of the plasma era into the domain of quark physics with large values of the baryonic chemical potential, where a number of remarkable phenomena may be observed experimentally in the near future, in facilities currently under construction. However, Little Inflation itself is compatible with values of $\mu_B/T$ well beyond those accessible to those facilities. It is therefore very interesting that, when holography is applied to this theory, one finds that $\mu_B/T$ is constrained to a very narrow range (the inequalities (\ref{W}) above); this much narrower range will indeed probably be reached by the experiments we mentioned.

If the Little Inflation picture is correct, then those experiments will be examining a system which (in some important aspects, though not all) closely resembles the early Universe during a brief but crucial period: the time when it was about to hadronize through a first-order phase transition. In short, we could soon be witnessing ``experimental early-Universe cosmology'' in a very non-trivial sense. (On the other hand, holography indicates that a still more remarkable possibility compatible with Little Inflation, that hadronization might take place near the quark matter \emph{triple point} \cite{kn:andronicover}, is very unlikely.)

If phenomena like chromodynamic critical opalescence are actually observed in these experiments, it will be important to consider whether such effects are compatible with established cosmological observations and theory, if the cosmic plasma passed very near to the quark matter critical point on its passage through the quark matter phase diagram. One may well find that these fluctuation phenomena, which can be quite dramatic, are ruled out by the observational data in the cosmic case. If so, the implication would be that the Little Inflation trajectory shown in Figure 1 stays well away from the critical point. As we have seen, holography implies that there is very little leeway for that, meaning that the plasma must have hadronized at the extreme upper end of the range given in (\ref{W}) above.

With a better understanding of the precise shape and slope of the phase line, one could use this to make a fairly precise prediction as to the location of the point in the phase diagram where the Universe hadronized; that is, one could predict the temperature and baryonic chemical potential at the beginning of the hadron era. Confirmation of such a prediction might be interpreted as strong evidence in favour of holography.

\addtocounter{section}{1}
\section*{\large{\textsf{Acknowledgements}}}
The author is grateful to Prof Soon Wanmei for technical assistance, and to Cate Yawen McInnes for encouraging him to complete this work expeditiously.

\addtocounter{section}{1}
\section*{\large{\textsf{Appendix: The Effect of a Magnetic Field}}}
As we explained above, Little Inflation provides a theory of cosmic magnetogenesis, but the magnetic fields involved are not enormously large; so we approximated $B$ by zero in the inequality (\ref{Q}). One should however consider the consequences if that should prove to be incorrect.

Since, in the conventional picture adopted here, $B$ and $T^2$ evolve in the same way during the plasma era, it is natural to set
\begin{equation}\label{X}
B\;=\;\beta \, T^2,
\end{equation}
where $\beta$ is a positive constant, the value of which will be considered below. It will be convenient also to express this parameter in a different way,
\begin{equation}\label{Y}
\alpha^2\;\equiv \;4\pi^3\;-\;\beta^2;
\end{equation}
it is clear from (\ref{Q}) that $\alpha$ can be assumed real and positive. Using this parameter, we can now express (\ref{Q}) as
\begin{equation}\label{Z}
r_h\;\leq \;{\alpha \,T^2\,L^2\over \mu_B}.
\end{equation}
Now the quartic on the left in equation (\ref{H}) is an increasing function at and beyond $r_h$, its largest real root, so substituting the right side of (\ref{Z}) into it we must obtain a non-negative expression. With some simple manipulations (in the course of which $L$ once again drops out) one then finds
\begin{equation}\label{ALPHA}
\left (4\pi^3 - {8\alpha^2\over 9}\right )\varsigma_B^4\;+\;\alpha^3\varsigma_B \;-\;{3\alpha^4\over 4\pi}\;\leq \;0.
\end{equation}
This is of course a quartic in $\varsigma_B$ of the same kind as the one in the inequality (\ref{U}); it reduces to the latter when $\beta = 0$. Again, therefore, $\varsigma_B$ is bounded above by the positive root. It is elementary to show that, if one thinks of this root as a function of $\alpha$, it is an increasing function: that is, it is a \emph{decreasing} function of $\beta$. Hence our claim that the inclusion of a magnetic field would only serve to strengthen our bound, inequality (\ref{V}). In practice, however, the extent of this strengthening is negligible, as we now show.

The largest value of $B/T^2$ considered in theories of cosmic baryogenesis arises \cite{kn:reviewB} when one considers the possibility of equipartition between the magnetic field energy density and the energy density of the plasma. We stress that such large values do \emph{not} normally arise in Little Inflation magnetogenesis, so the situation we are considering now is very much an over-estimate of the effect. In any case, the Stefan-Boltzmann law implies that, at equipartition,
\begin{equation}\label{BETA}
B\;\approx\; \sqrt{{2\over 15}}\pi T^2;
\end{equation}
this corresponds to about $3.7 \times 10^{17}$ gauss at the phase transition: but it only translates to $\beta \approx 1.15$. Computing the corresponding value of $\alpha$, inserting it into the left side of (\ref{ALPHA}), and solving numerically, one obtains
\begin{equation}\label{GAMMA}
\mu_B/T \; \leq \;\approx 2.324.
\end{equation}
Comparing this with (\ref{V}), one sees that, even in the most extreme case, the inclusion of a magnetic field does not materially affect our conclusions.

\end{document}